\newcommand{\dd}{{\rm d}}
\documentclass[10pts,showpacs,preprint,aps]{revtex4}
\usepackage{graphicx}% Include figure files
\usepackage{dcolumn}% Align table columns on decimal point
\usepackage{bm}% bold math
\usepackage{amssymb}
\usepackage{amsmath}
\begin{document}
\setcounter{page}{1}
\vskip 2cm
\title
{Velocity and velocity bounds in static spherically symmetric metrics}
\author
{I. Arraut$^1$, D. Batic$^{2,3}$ and M. Nowakowski$^4$}
\affiliation{$^1$
Centro de Investigaciones, Universidad Antonio Nari\~no,
Cra.3 Este No.47A-15, Bogot\'a, Colombia\\
$^2$ Department of Mathematics, University of the West Indies,
Kingston 7, Jamaica\\
$^3$ Departamento de Matem\'aticas, Universidad de los Andes,
Cra 1E, No. 18A-10, Bogot\'a, Colombia\\
$^4$
Departamento de F\'{\i}sica, Universidad de los Andes,
Cra.1E No.18A-10, Bogot\'a, Colombia
}

\begin{abstract}
We find simple expressions for velocity of massless particles in dependence of the distance $r$
in Schwarzschild coordinates.
For massive particles these expressions put an upper bound
for the velocity.
Our results apply to static spherically symmetric metrics.
We use these results to calculate the velocity for different cases: Schwarzschild, Schwarzschild-de Sitter and
Reissner-Nordstr\"om with and without the cosmological constant. We emphasize the differences between
the behavior of the velocity in the different metrics and find that in cases with naked
singularity there
exists always a region where the massless particle moves with a velocity bigger than the velocity of light in vacuum.
In the case of Reissner-Nordstr\"om-de Sitter we completely characterize the radial velocity and the metric
in an algebraic way. We contrast the case of classical naked singularities with naked singularities
emerging from  metric inspired by noncommutative geometry where the radial velocity never exceeds one.
Furthermore, we solve the Einstein equations for a constant and polytropic density profile
and calculate the radial velocity of a photon moving in spaces with interior metric.
The polytropic case of radial velocity displays an unexpected variation bounded by a local minimum and
maximum.
\end{abstract}
\pacs{04.25.Nx, 04.20.Dw, 04.70.Bw, 95.30.Sf}
\maketitle
\section{Introduction}
The class of static spherically symmetric metrics is widely used
in General Relativity and Astrophysics \cite{Carroll,Weinberg}.
The trajectory of a test body in the gravitational field
determined by this metric can be obtained in a closed form up to
an integral. The latter is not always solvable in a compact form.
Therefore, more direct analytical and easily accessible
information on the motion of the test body is welcome. One such
possible method is to cast the equation of motion into a form
resembling the corresponding expression of classical mechanics
with an effective potential $U_{\rm eff}$. This is certainly
possible for exterior metrics and the power of this method for the
Schwarzschild-de Sitter case (with a cosmological constant
$\varLambda$) has been demonstrated in \cite{marek}. Here, we
attempt a yet different way to gain information about the particle
motion by examining the radial and angular velocity of the
particle in dependence of the distance. Such an information can be
derived without the necessity of solving the full equation of
motion. The usefulness of the results can be then shown by
applying the general result to the different cases of exterior and
interior spherically symmetric metrics. For instance, in contrast
to the Schwarzschild metric, where the radial velocity of a
massless particle approaches asymptotically one at large
distances, the same quantity in the Schwarzschild-de Sitter metric
reaches a maximal value $1-(3r_s r_{\varLambda}^{-1})^{2/3}$ at
$r=r_1\equiv (3r_s r_{\varLambda}^2)^{1/3}$
($r_{\varLambda}=1/\sqrt{\varLambda}$) beyond which it decreases.
Note that $r_1$ is of astrophysical order of magnitude as it is a
combination of a small quantity, $r_s=GM$, with a large one
$r_{\varLambda}$. Indeed, $r_1$ is also of the order of magnitude
of the typical extension of a  star cluster (if the mass $M$ is a
typical star mass), galaxy (if $M$ is of the order of magnitude of
a star cluster) and galaxy cluster (choosing $M$ to be the mass of
a typical galaxy)  \cite{marek}. As we will show, this implies the
interesting result of the Schwarzschild-de Sitter space-time, i.e.
the velocity of a photon leaving one of the above astrophysical
objects does not grow with distance (after having travelled a
distance $r_1$) to approach asymptotically one, but decreases
monotonically. The velocity expressions which we derive allow also
a deeper understanding of the special role of naked singularities.
As we will demonstrate in the case of naked singularities, the
radial velocity of a massless particle exceeds in the
Schwarzschild coordinates in certain space regions the value one,
i.e. the velocity of light in vacuum (and in the absence of
gravitational fields). The naked singularities which we study in
this work are of Reissner-Nordstr\"om and Reissner-Nordstr\"om- de
Sitter type. In the latter case we give an algebraic
characterization of the metric and the radial velocity of a
massless particle deriving among other the conditions for the
existence of the naked singularity and for the regions where the
radial velocity is bigger than one. We give a definition of the
extreme Reissner-Nordstr\"om metric in terms of the radial
velocity.
We also study the case of naked singularities encountered in metrics
inspired by noncommutative geometry. We find that in contrast
to the classical case the radial velocity of a massless
particles is always smaller-equal one.
We apply our velocity expressions also to the case of an
interior metric. We do this first for the case of constant density
and later assume the density to be determined by a polytropic
equation of state and the Tolman-Oppenheimer-Volkov hydrostatic
equilibrium equation. In both cases, the radial velocity of the
photon will be smaller than one and approximately one if the
gravitational field is weak. The interesting feature is, however,
that in the polytropic case it will display a variation with some
local minima and maxima present.

The concept of a velocity i.e. the change of distance with respect
to the time coordinate is in General Relativity a frame dependent
concept as it is not an invariant. This is not a drawback as it
corresponds to the real measurements we perform in practice.
Although the Equivalence Principle assures that we can always find
a local frame in which the velocity of a photon is the velocity of
light (as defined in vacuum), this is not the frame which is
conform with our measurement devices. Indeed, all our measurements
of local (as opposed to cosmological) effects of General
Relativity refer explicitly to Schwarzschild coordinates in which
we establish and measure the precession of perihelia, deflection
of light or radar echo delay. In these coordinates the radial
velocity of light is not a constant as it is determined by the
geodesic equation of motion. It is therefore of some interest to
study the radial velocity of a photon in Schwarzschild
coordinates.

The paper is organized as follows. In section 2 and 3 we derive the
general expression for the radial velocity in Schwarzschild coordinates in dependence
of the radial coordinate. In section 4 we apply the radial velocity expressions to
vacuum metric and discuss the significance of the results. In section 5 we study
the radial velocity of a massless particle in interior metric using constant density
and a polytropic model. In section 6 we present the conclusions.

\section{The general static spherically symmetric metric}   \label{eq:s1}
Let us remind the reader that the most general form of a static
spherically symmetric metric can be written as
\begin{equation}\label{eq:3}
\dd s^2=-A(r)\dd t^2+B(r)\dd r^2+r^2(\dd\vartheta^2+\sin^2{\vartheta}\dd\varphi^2).
\end{equation}
As it is well known there exist two associated conserved
quantities \cite{Carroll,Weinberg}
\begin{equation}   \label{eq:6}
E=A(r)\frac{dt}{d\tau}
\end{equation}
and
\begin{equation}   \label{eq:7}
r_l=r^2\frac{\dd\phi}{\dd\tau}=r^2\dot{\phi}\frac{E}{A(r)},
\end{equation}
where $\tau$ is the proper time, i.e. we have chosen the affine
parameter to be $\tau$ and dotted quantities signify derivative
with respect to time $t$. Another constant which emerges due to
the geodesic equation taken together with the metric compatibility
is
\[
\epsilon=-g_{\mu \nu}\frac{\dd x^\mu}{\dd \tau}\frac{\dd x^\nu}{\dd\tau},
\]
where $\epsilon=1$ for massive particles and $\epsilon=0$ for
massless particles. Rewriting the above expression by means of the
metric (\ref{eq:3}) we obtain
\[
-\frac{\epsilon}{B(r)}=-\frac{A(r)}{B(r)}\left(\frac{\dd t}{\dd\tau}\right)^2
+\left(\frac{\dd r}{\dd\tau}\right)^2+\frac{r^2}{B(r)}\left(\frac{\dd\vartheta}{\dd\tau}\right)^2+
\frac{r^2\sin^2{\vartheta}}{B(r)}\left(\frac{\dd\varphi}{\dd\tau}\right)^2.
\]
The conserved quantities given in (\ref{eq:6}) and (\ref{eq:7})
enable us to cast the above equation into the form
\begin{equation}   \label{eq:12}
\frac{1}{2}\left(\frac{\dd
r}{\dd\tau}\right)^2+\frac{r_l^2}{2r^2}\frac{1}{B(r)}
=\frac{E^2}{2A(r)B(r)}-\frac{\epsilon}{2B(r)},
\end{equation}
which is essentially the equation of motion where the plane of
motion is such that $\theta=\pi/2$. Taking into account that for
the metric (\ref{eq:3}) we have $A(r)B(r)=1$, equation
(\ref{eq:12}) becomes
\begin{equation} \label{eq:12ex}
\frac{1}{2}\left(\frac{\dd r}{\dd\tau}\right)^2 +U_{\rm eff}=\frac{E^2}{2}, \,\,\, U_{\rm eff}=
\left(\frac{r_l^2}{2r^2}+\frac{\epsilon}{2}\right)A(r).
\end{equation}
Instead of solving directly (\ref{eq:12}) or (\ref{eq:12ex}), we
will pursue here a different strategy. First of all, notice that
the conserved quantities (\ref{eq:6}) and (\ref{eq:7}) permit us
to write equation (\ref{eq:12}) as
\begin{equation}   \label{eq:15}
E^2 \left(1-\frac{\dot{r}^2}{A^2(r)}
-\frac{r^2\dot{\phi}^2}{A(r)}\right)=\frac{\epsilon}{B(r)}.
\end{equation}
Similarly, we can express (\ref{eq:7}) in the following form
\begin{equation}   \label{eq:16}
r_l^2\left(\frac{1}{B^2(r)}-\dot{r}^2-\frac{r^2\dot{\phi}^2}{B(r)}\right)=r^4\dot{\phi}^2\frac{\epsilon}{B(r)}.
\end{equation}
In the subsequent sections the last two equations will be applied
to different cases of exterior and interior spherically symmetric
metrics.

\section{Velocity and velocity bounds}   \label{eq:s2}
We assume first $\epsilon =1$, i.e. the case of massive particles.
In all examples we will discuss later the metric elements $B(r)$
and $A(r)$ satisfy $B(r) \ge 0$ and $A(r) \ge 0$ for $r\in
K\subset\mathbb{R}^{+}$, where the explicit form of $K$ depends on
the particular metric under consideration. For instance, in the
Schwarzschild case we would have $K=[2r_s,\infty)$. Let $v_r(r)$
and $v_\phi(r)$ be the radial and tangential velocities of a test
particle, respectively. Then, for $\epsilon=1$ the r.h.s. of
equation (\ref{eq:15}) must be positive definite and we obtain
\[
1-\frac{v^2_r(r)}{A^2(r)}-\frac{v^2_\phi(r)}{A(r)}>0.
\]
This expression gives us the velocity bounds of a massive
particle. In particular, notice that if the motion is purely
radial, i.e. $\dot{\phi}=0$, then we obtain the following bound
for the radial velocity
\[
v_r(r)< A(r),\quad r\in K\subset\mathbb{R}^{+},
\]
whereas in the case of purely tangential motion, that is
$\dot{r}=0$, a similar bound can be found, namely
\[
v_\phi(r) < \sqrt{A(r)},\quad r\in K\subset\mathbb{R}^{+}.
\]
In the massless case the inequalities we derived for massive
particles become equalities and we have
\begin{equation}   \label{eq:20x}
v_r(r)=A(r),\quad v_\phi(r)= \sqrt{A(r)}.
\end{equation}
Notice that for the Minkowski metric in spherical coordinates $\dd
s^2=-\dd t^2+\dd r^2+r^2\dd\Omega^2$ where $A(r)=B(r)=1$, we
obtain as expected $v_r(r) < 1 $ for the massive case and
$v_r(r)=1$ for the photon's radial velocity.

\section{Vacuum metrics}   \label{eq:s4}
In this section we discuss the velocity of a test particle in
dependence of the distance $r$ for different types of metrics. The
results are depicted in figures 1-4, which display the main
features of the radial velocity as zeros, local extrema and
regions where the velocity is bigger than one. In the figures we
have not used the actual values of $r_s=G_NM$ and
$r_\varLambda=1/\sqrt{\varLambda}$ (where $G_N$ is the Newtonian
and $\varLambda$ the cosmological constant) as both these scales
are very far apart. We therefore decided to use arbitrary units in
the figures. Notice that a common feature of the velocity profiles
is that $v_r(r)$ vanishes when evaluated at the event horizon
$r_H$ since light cannot escape from a black hole as seen from the
standpoint of a distant observer in Schwarzschild coordinates.

\subsection{Schwarzschild and Schwarzschild-de Sitter metric}
\begin{figure}
\begin{center}
\scalebox{.5}{\includegraphics{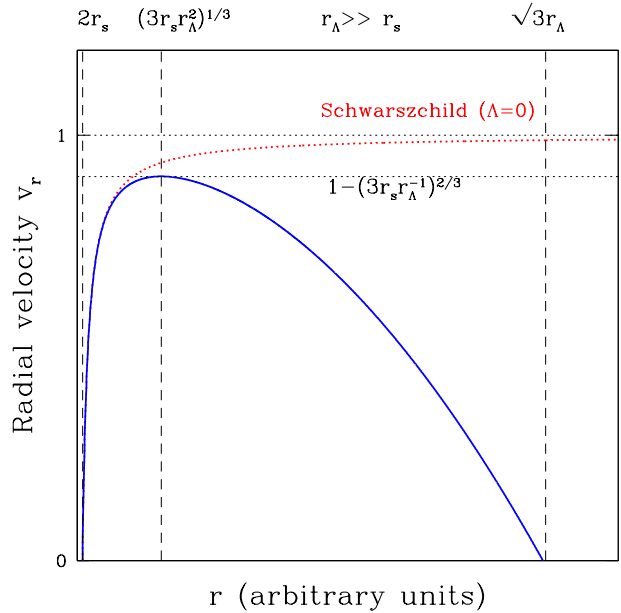}}
\caption{Comparison of the radial velocities of photon moving in
the Schwarzschild and Schwarzschild-de Sitter metric. Evidently,
the photon velocity does not reach $1$, not even asymptotically,
in the latter case and decreases after a maximum value smaller
than one.}
\end{center}
\end{figure}
In the case of Schwarzschild-de Sitter, also called Kottler metric
\cite{Kottler}
\[
A(r)= 1-\frac{2r_s}{r}-\frac{1}{3}\frac{r^2}{r_\varLambda^2}
\]
the velocities are given by
\[
v_r(r)=1-\frac{2r_s}{r}-\frac{1}{3}\frac{r^2}{r_\varLambda^2},\quad
v_\phi(r)=
\sqrt{1-\frac{2r_s}{r}-\frac{1}{3}\frac{r^2}{r_\varLambda^2}},
\]
where the Schwarzschild case is recovered for $r_{\varLambda} \to
\infty$. The interesting feature emerging from FIG. 1 is that in
the Schwarzschild-de Sitter metric the photon radial velocity
reaches a maximum velocity smaller than one at a finite distance
$r_1=(3r_sr_{\varLambda}^2)^{1/3}$ where
\begin{equation} \label{deSitter}
v_r(r_1)=1- \left(\frac{3r_s}{r_{\varLambda}}\right)^{2/3}
\end{equation}
after which $v_r$ decreases. Moreover, from (\ref{deSitter}) we
also see that in order there exists an interval of the radial
variable $r$ such that $A(r)>0$ it must be
\begin{equation}\label{cond_I}
r_\Lambda>3r_s.
\end{equation}
Under condition (\ref{cond_I}) the radial velocity will become
zero at $\widetilde{r}_0\approx 2r_s$ and
$r_c\approx\sqrt{3}r_\Lambda$. Hence, we have
\[
0<v_r(r)<v_r(r_1)<1\quad\mbox{for}\quad
r\in(\widetilde{r}_0,r_c)\quad\mbox{and}\quad r_\Lambda>3r_s.
\]
Note that $r_1$ is a combination of a small parameter $r_s$ with a
large one, i.e. $r_{\varLambda}$. Therefore, its actual value is
of astrophysical order of magnitude even if $r_{\varLambda}$ is
very large. The actual value of $r_1$ can be estimated by using
the currently favored value $\rho_{\rm vac} \simeq 0.7 \rho_{\rm
crit}$ together with
\[
\Lambda=3\left(\frac{\rho_{vac}}{\rho_{crit}}\right)H_0^2
\]
where $H_0$ is the Hubble constant. For sun masses it comes out to
be approximately the size of a globular cluster, for the mass of a
globular cluster it is roughly the size of a average galaxy and
finally, inserting the mass of a galaxy $r_1$ approaches the size
of a galaxy cluster \cite{marek}.  The distance $r_1$ has a
fourfold meaning
\begin{itemize}
\item[(i)]
It is the maximal size of a bound orbit in Schwarzschild-de Sitter
metric in the approximation of small angular momentum
\cite{marek}.
\item[(ii)]
It is the maximal size of virialized matter in space-time with
$\varLambda$ \cite{marek}.
\item[(iii)]
As shown above, it is the distance at which a photon reaches its
maximal velocity smaller than one. This curious features of the
photon radial velocity can have an astrophysical effect, namely
the radial velocity of a photon leaving the surface of a typical
galaxy and after travelling roughly a distance of about the size
of a galaxy cluster starts decreasing with distance.
\item[(iv)] It is the critical distance after which the radiation is blue-shifted \cite{we}.
\end{itemize}

It is also clear that the velocity of light as measured in
Schwarzschild coordinates at the earth's surface is not the
velocity of light which we would measure in an empty space
(Minkowski space-time). Measured in the radial direction the
velocity of light on earth would be $1-1.39 \times 10^{-9}$ which
has a small historical relevance. In the second half of the twenty
century it was thought that a more accurate measurement of the
velocity of light was hampered by inaccurate definition of the
meter \cite{Evenson}. Therefore in 1983 the meter has been
redefined as the length travelled by light in vacuum during a time
interval of $1/299792458$ seconds \cite{BIPM}. This definition
makes the velocity of light in vacuum an exact number:
$299792458$m/s. The best value of directly measured velocity of
light is $299792456.2\pm 1.1$m/s \cite{Evenson}. This means that a
direct measurement achieves an accuracy of about $10^{-9}$ which
is exactly of the same size as the effect of measuring the radial
velocity of light at the surface of earth. This also means that
such an effect is actually not (yet) measurable. One could object
that our estimate is based on a constant velocity concept whereas
the actual radial velocity in Schwarzschild coordinates depends on
$r$. By defining an average
\begin{equation} \label{average}
<v_r>=\frac{1}{r_b -r_a}\int_{r_a}^{r_b}v_r(r)dr
\end{equation}
and choosing $r_a=R_{\bigoplus}$, i.e. the average radius of the
earth  and $r_b=R_{\bigoplus}-\delta r$, we obtain
\begin{equation} \label{average2}
<v_r>=1-\frac{2r_s}{\delta r}\ln\left(1 + \frac{\delta r}{R_{\bigoplus}}\right)\simeq 1 -
\frac{2r_s}{R_{\bigoplus}}+\frac{2r_s}{R_{\bigoplus}}\frac{\delta r}{R_{\bigoplus}}
\end{equation}
which shows that one can neglect the $\delta r$ corrections near the earth surface.

\subsection{Noncommutative inspired Schwarzschild and Reissner-Nordstr\"om metrics}
An interesting class of static, spherically symmetric metrics
inspired by noncommutative geometry  consists of the so called
noncommutative geometry inspired Schwarzschild and
Reissner-Nordstr\"om metrics \cite{NCSM}. The noncommutative
geometry inspired Schwarzschild black hole has line element (in
this subsection we choose $c=G=1$)
\begin{equation}\label{NCSM-metric}
ds^2=g(r)dt^2-g(r)^{-1}dr^2-r^2(d\vartheta^2+\sin^2{\vartheta}d\varphi^2)
\end{equation}
where
\[
g(r)=1-\frac{2M_S(r)}{r},\quad M_S(r)=
\frac{2M}{\sqrt{\pi}}\gamma(3/2,r^2/4\theta),\quad
\gamma(3/2,r^2/4\theta)=\int_0^{r^2/4\theta}d\tau~\sqrt{\tau}e^{-\tau}
\]
and $\theta$ is a parameter encoding noncommutativity. Notice that
for $r/\sqrt{\theta}\to\infty$ the line element
(\ref{NCSM-metric}) becomes the classical Schwarzschild metric.
For $M<M_0=1.9\sqrt{\theta}$ (\ref{NCSM-metric}) (we assume that
at least formally it is possible to have such an $M$) does not
exhibit any horizon and describes what we call a naked,
self-gravitating droplet of anisotropic fluid \cite{NCSM}.
Finally, for $M>M_0$ there exist two horizons and only one horizon
at $r_0=3.0\sqrt{\theta}$ for $M=M_0$. The noncommutative geometry
inspired Schwarzschild metric has a line element of the form
\begin{equation}\label{NCRN-metric}
ds^2=f(r)dt^2-
f(r)^{-1}dr^2-r^2(d\vartheta^2+\sin^2{\vartheta}d\varphi^2)
\end{equation}
with
\begin{eqnarray*}
f(r)&=&1-\frac{4M}{r\sqrt{\pi}}\gamma(3/2,r^2/4\theta)+\frac{Q^2}{\pi
r^2}\left[F(r)+\sqrt{\frac{2}{\theta}}r\gamma(3/2,r^2/4\theta)\right],\\
F(r)&=& \gamma^2(1/2,r^2/4\theta)-\frac{r}{\sqrt{2\theta}}\gamma(1/2,r^2/2\theta),\quad
\gamma(a/b,x)=\int_{0}^x\frac{du}{u}~u^{a/b}e^{-u},\\
\end{eqnarray*}
where $M$ and $Q$ are the mass and charge parameters,
respectively. Asymptotically for $r/\sqrt{\theta}\to\infty$ the
above line element reproduces the classic Reissner-Nordstr\"om
metric. Depending on the values of both $M$ and $Q$ the line
element (\ref{NCRN-metric}) describes a charged black hole with
two horizons, an extremal charged black hole or a naked, charged,
self-gravitating droplet of anisotropic fluid. We do not give the
algebraic details here to define the naked singularity in the
charged case, but rely on numerical results. Since the above
metrics are static and spherically symmetric we can apply our
formula (\ref{eq:20x}) to study the radial velocity of a massless
particle in the aforementioned metrics. In Figure 2 we plot the
radial velocity of a photon moving in the above metrics under the
condition that we have a naked droplet. It is evident that a
massless particle starting with velocity one asymptotically away
from the droplet will experience a deceleration as it approaches
the droplet till it reaches a minimum velocity at
$r_m=3\sqrt{\theta}$. For $r<r_m$ the particle starts accelerating
and its velocity will be one again at $r=0$. Worth noticing is the
fact that the velocity is everywhere smaller-equal one. This is
different if the naked singularity were of classical type, a case
discussed in the next subsection.
\begin{figure}
\begin{center}
\scalebox{.5}{\includegraphics{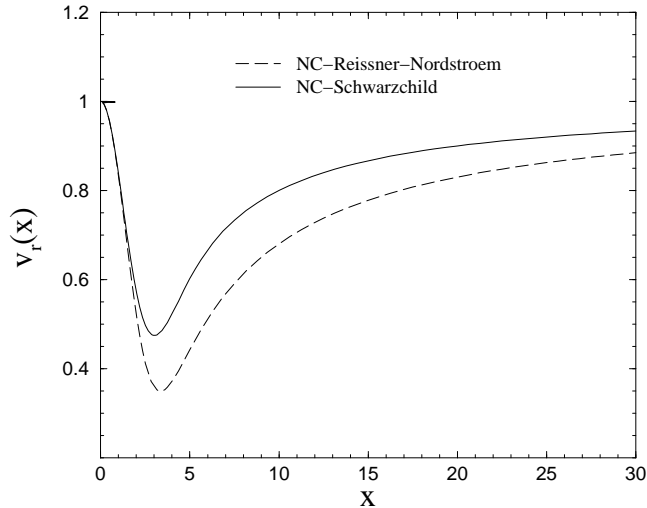}}
\caption{\footnotesize{Photon radial velocity versus
$x=r/\sqrt{\theta}$ in metrics inspired by noncommutative geometry
with a naked singularity. The upper curve corresponds to the
Schwarzschild noncommutative case whereas the lower one refers to
noncommutative Reissner-Nordstr\"om. The choice of the parameters
is: $M/\sqrt{\theta}=1$, $Q/\sqrt{\theta}=2$. The velocity is
always smaller-equal one.}}
\end{center}
\end{figure}

\subsection{What is indecent about naked singularities?}
\begin{figure}
\begin{center}
\scalebox{.5}{\includegraphics{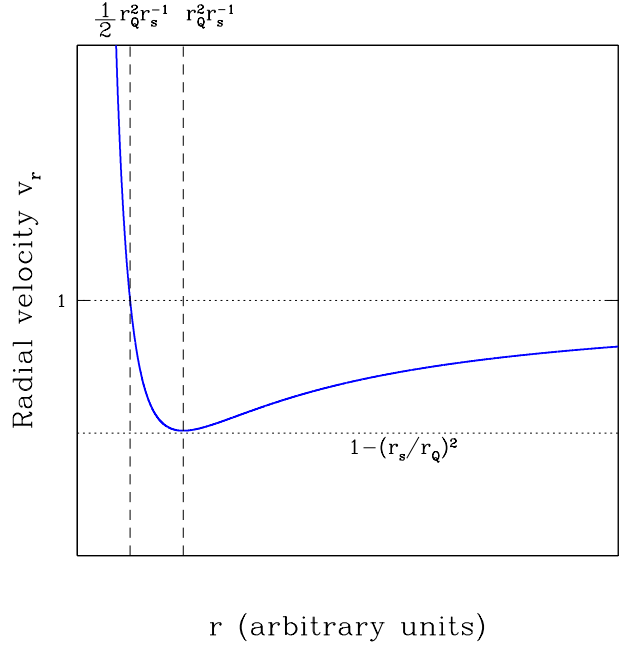}}
\caption{\footnotesize{Photon radial velocity in a
Reissner-Nordstr\"om black hole with naked singularity condition
$r_Q > r_s$. There exists a region where the velocity exceeds
one.}}
\end{center}
\end{figure}
We will discuss here the effect of naked singularity on the radial
velocity of a photon in the Reissner-Nordstr\"om metric \cite{RN}
with and without cosmological constant. In the present case the
function $A(r)$ is given by
\begin{equation}\label{profilo}
A(r)=1-\frac{2r_s}{r}+\frac{r_Q^2}{r^2}-\frac{1}{3}\frac{r^2}{r_\varLambda^2}.
\end{equation}
This allows us to calculate the velocities as follows
\[
v_r(r)=
1-\frac{2r_s}{r}+\frac{r_Q^2}{r^2}-\frac{1}{3}\frac{r^2}{r_\varLambda^2},\quad
v_\phi(r)=
\sqrt{1-\frac{2r_s}{r}+\frac{r_Q^2}{r^2}-\frac{1}{3}\frac{r^2}{r_\varLambda^2}}.
\]
The Reissner-Nordstr\"om case is obtained in the limit
$r_{\varLambda} \to \infty$. Notice that in this limit and for
$r_s<r_Q$ the radial velocity exceeds the velocity of light
(represented by the value $1$) below
\begin{equation} \label{RN}
r_0=\frac{1}{2}\frac{r^2_Q}{r_s},
\end{equation}
and displays a local minimum at
\[
r_1=\frac{r^2_Q}{r_s},\quad
v_r(r_1)=1-\left(\frac{r_s}{r_Q}\right)^2,
\]
after which it approaches one. A similar behavior can be seen if a
positive cosmological constant is introduced. To this purpose it
is interesting to study under which conditions on the parameter
$r_s$, $r_Q$ and $r_\varLambda$ the Reissner-Nordstr\"om-deSitter
geometry exhibits a naked singularity. In what follows the only
hypothesis we will make is that $r_\varLambda>r_s$ and
$r_\varLambda>r_Q$. The minimal requirement for the emergence of a
naked singularity is that (\ref{profilo}) possesses two real
intersections with the $r$-axis having opposite signs. This
condition assures the only existing horizon is the cosmological
one. Finally, there is a second characteristic point important to
define the extreme case, namely that the radial velocity has local
minimum $r_m$ and one local maximum $r_M$. Concerning the first
requirement, we have to study the roots of the equation
\begin{equation}\label{stella2}
-\frac{r^4}{3r^2_\varLambda}+r^2-2r_s r+r^2_Q=0.
\end{equation}
According to \cite{Cohen} the above quartic equation will have two
real roots and one conjugate pair of complex roots if its
associated discriminant $\Delta_I$ is negative, that is
\[
\Delta_I=4I_{I}^3-J^2_{I}<0
\]
where
\[
I_I=1-4~\frac{r^2_Q}{r^2_\varLambda},\quad
J_I=-24~\frac{r^2_Q}{r^2_\varLambda}+36~\frac{r_s^2}{r_\varLambda^2}-2.
\]
Hence, the function $v_r(r)$ will intersect the $r$-axis at two
different locations if
\begin{equation}\label{inequality1}
(r_\varLambda^2-4r_Q^2)^3-r_\varLambda^2(12
r^2_Q-18r_S^2+r_\varLambda^2)^2<0.
\end{equation}
Fig.~3 shows that in the naked singularity scenario the radial
velocity function may exhibit one local minimum and maximum or a
saddle point. In order that $v_r(r)$ has at least a local minimum
and maximum, we have to require that $dv_r/dr=0$. This leads to
the quartic
\begin{equation}\label{quart_vel}
-\frac{r^4}{3r^2_\varLambda}+r_s r-r^2_Q=0.
\end{equation}
The above equation will have two real roots and one conjugate pair
of complex roots if the discriminant $\Delta_{II}$ is negative,
that is
\[
\Delta_{II}=4I_{II}^3-J^2_{II}<0
\]
with
\[
I_{II}=4~\frac{r^2_Q}{r^2_\varLambda},\quad J_{II}=
9~\frac{r^2_s}{r^2_\varLambda}.
\]
Therefore, the condition for the existence of a local minimum and
maximum reads
\begin{equation}\label{inequality2}
r_Q<r_Q^{crit},\quad
r_Q^{crit}=\left(\frac{3}{4}\right)^{2/3}(r_\varLambda r_s^2)^{1/3}.
\end{equation}
Moreover, the real solutions of (\ref{quart_vel}) are
\[
r_{m,M}=\frac{\sqrt{2}}{4}\left(\alpha^{1/2}\mp
\sqrt{12\sqrt{2}~\frac{r_s
r_\varLambda^2}{\alpha^{1/2}}-16~\frac{r^2_Q
r_\varLambda^2}{\beta^{1/3}}-\beta^{1/3}}\right)
\]
with
\[
\alpha=\frac{\beta^{2/3}+16~r_Q^2
r_\varLambda^2}{\beta^{1/3}},\quad\beta=36~r_s^2
r_\varLambda^4+4\sqrt{-256~r_Q^6 r_\varLambda^6+81~r_s^4 r_\varLambda^8}.
\]
Notice that (\ref{inequality2}) ensures that $\beta>0$. Moreover,
there are two sign changes in (\ref{quart_vel}), hence Descartes
sign rule implies that the maximum number of positive roots is
two. On the other hand, the polynomial equation obtained from
(\ref{quart_vel}) by reversing the sign of $r$ does not exhibit
any sign change and we conclude that the maximum number of
negative real roots must be zero. From these considerations it
follows that $r_m>0$ whenever (\ref{inequality2}) is satisfied.
Finally, the radial velocity evaluated at the local minimum will
be always positive in virtue of (\ref{inequality1}). Furthermore,
we can characterize an extreme Reissner-Nordstr\"om-de Sitter
black hole by a choice of the parameters $r_s$, $r_Q$ and
$r_\varLambda$ such that $v_r(r_m)=0$ together with the condition
that (\ref{stella2}) has four real roots with two of them
coinciding. In other words, this will ensure the existence of only
one extra horizon apart from the cosmological one. For the sake of
brevity, we will treat here only the special case
$r_{\varLambda}r_s^2 \gg r_Q^3$ implying $r_Q \ll r_Q^{crit}$. In
terms of the small parameter $\gamma=r_Q/r_Q^{crit}$ the local
minimum is located at
\begin{equation} \label{min}
r_m=\left(\frac{3}{4}\right)^{4/3} (r_{\varLambda}^2 r_s)^{1/3}
\gamma^2\left[1+\frac{27}{256}\gamma^6+\mathcal{O}(\gamma^{12})\right]
\end{equation}
whereas the radial velocity at $r_m$ takes the value
\[
v_r(r_m)=1-\left(\frac{4}{3}\right)^{4/3}\left(\frac{r_s}{r_{\varLambda}}\right)^{2/3}\frac{1}{\gamma^2}
+\mathcal{O}(\gamma^4).
\]
Notice that in physical interesting cases $r_s\ll r_\varLambda$.
Hence, the quantity $(r_s/r_\varLambda)^{2/3}\gamma^{-2}$ will be the
product of a small quantity multiplying a large one. A simple
computation reveals that $(r_s/r_\varLambda)^{2/3}\gamma^{-2}\propto
(r_s/r_Q)^2$ and we conclude that
$(r_s/r_\varLambda)^{2/3}\gamma^{-2}$ will be of order one if
$r_s\approx r_Q$. Finally, we find the maximum at $r_M$
\begin{equation} \label{max}
r_M=(3r_{\varLambda}^2
r_s)^{1/3}\left[1-\frac{\gamma^2}{2^{8/3}}-\frac{\gamma^4}{2^{13/3}}+\mathcal{O}(\gamma^6)\right]
\end{equation}
with
\begin{equation} \label{max2}
v_r(r_M)=1-\left(3\frac{r_s}{r_{\varLambda}}\right)^{2/3}+
\frac{1}{2^{8/3}}\left(3\frac{r_s}{r_{\varLambda}}\right)^{2/3}\gamma^2+\mathcal{O}(\gamma^4).
\end{equation}
It is not difficult to see that at the fourth order in $\gamma$ we
always have $v_r(r_M)<1$. The radial velocity becomes bigger than
one below a value given approximately by $r_0$ in equation
(\ref{RN}). In general, $v_r(r)<1$ for those values of $r$ such
that
\[
p(r)=-\frac{r^4}{3r^2_\varLambda}-2r_s r+r^2_Q>0.
\]
It is not difficult to check that the discriminant of the quartic
equation $p(r)=0$ is always negative, thus implying that there
will be always two real roots and one conjugate pair of complex
roots. Moreover, according to the Descartes sign rule there will
be one negative and one positive real root. The latter is
\[
R_0=\frac{1}{2}\left(-{\widetilde{\alpha}}^{1/2}+
\sqrt{12~\frac{r_s
r_\varLambda^2}{{\widetilde{\alpha}}^{1/2}}+4~\frac{r^2_Q
r_\varLambda^2}{{\widetilde{\beta}}^{1/3}}-{\widetilde{\beta}}^{1/3}}\right)
\]
with
\[
\widetilde{\alpha}=\frac{{\widetilde{\beta}}^{2/3}-4~r_Q^2
r_\varLambda^2}{{\widetilde{\beta}}^{1/3}},\quad\widetilde{\beta}=18~r_s^2
r_\varLambda^4+2\sqrt{16~r_Q^6 r_\varLambda^6+81~r_s^4 r_\varLambda^8}.
\]
If $r_{\varLambda}r_s^2 \gg r_Q^3$ then the critical radius below
which we encounter velocities bigger than one is
\[
R_0=\frac{1}{2}\left(\frac{3}{4}\right)^{4/3} (r_{\varLambda}^2
r_s)^{1/3}
\gamma^2\left[1-\frac{27}{4096}\gamma^6+\mathcal{O}(\gamma^{12})\right].
\]
Notice that at the second order in $\gamma$ we have $R_0\approx
r_m/2$ and in the regime $r_Q\ll r_Q^{\rm crit}$ the radial
velocity exceeds one whenever $r<r_m/2$. The question 'what is so
special about naked singularities' can be answered, using the
formulae we derived for the velocity, fast and efficiently. There
exist regions where the velocity of a massless particle is bigger
than the velocity of light as defined in special theory of
relativity. It is known that naked singularities are gravitational
singularities which are not hidden by an event horizon, and hence
they might be observable. If loop quantum gravity is correct, then
naked singularities could exist in nature, implying that the
cosmic censorship hypothesis does not hold. Numerical calculations
and some other arguments have also hinted at this possibility
\cite{naked}. Furthermore, it cannot be excluded that closed
time-like curves might arise in the vicinity of naked
singularities. For example in \cite{deFelice} the possibility of
strong causality violation in the presence of a naked singularity
has been analyzed. Moreover, for Kerr black holes with
sufficiently high rotational parameter the space-time contains a
naked singularity and also closed time-like curves. However, the
connection between naked singularity
and causality violation is not clear and deserves further studies \cite{Joshi}.\\
\begin{figure}
\begin{center}
\scalebox{.5}{\includegraphics{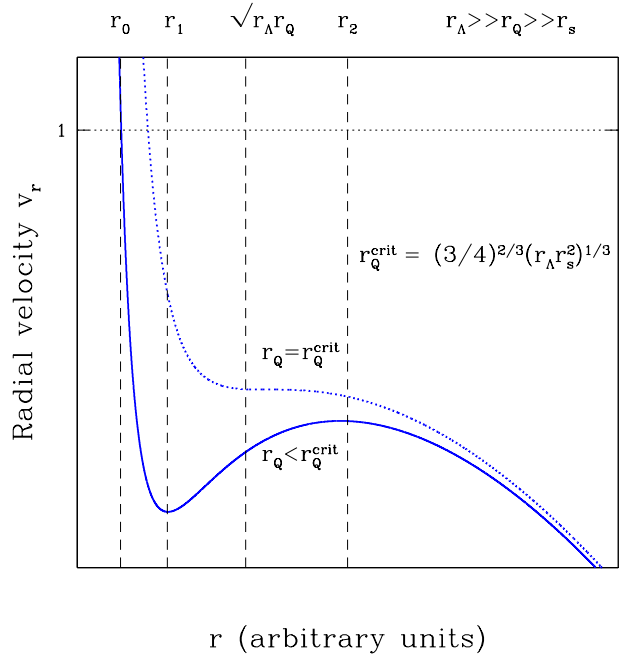}} \caption{The
radial velocity of a massless particle in a
Reissner-Nordstr\"om-de Sitter metric with a naked singularity.
The only (cosmological) horizon is at $\sqrt{3}r_{\varLambda} -r_s$. There exists
a region where the velocity exceeds one. For the explanations of
$r_0=R_0$, $r_1=r_m$ and $r_2=r_M$ see the text.}
\end{center}
\end{figure}
Two comments are in order here: we emphasize once again that as
long as we have the Equivalence Principle we can find a frame in
which locally the velocity of light is one. It is only in the
Schwarzschild coordinates that in the case of naked singularities
this velocity exceeds the value one. Should the Equivalence
Principle be broken (say, by some quantum effects) such a local
frame will not exist \cite{Shore}. Secondly, the aspects of the
particle motion in metrics with a cosmological constant can be
considered as examples of local (as opposed to cosmological)
effects of this constant. This is expected since the cosmological
constant is part of the Einstein tensor and not of the
energy-momentum tensor and will therefore affect, in principle,
any measurable quantity (for other local effects see
\cite{Lambda1, Lambda2, Lambda3}).
\section{Interior metrics}
For the interior metrics we have a density profile (e.g. a constant density or any other
form), say of the general form
\[
\rho(r)= \rho_0(r),\quad r \le R\quad\mbox{and}\quad
\rho(r)=0,\quad r \ge R
\]
for some $R$ which denotes the extension of the object, i.e. $P(R)=0$ with $P$ being the pressure
in the perfect fluid energy momentum tensor. We demand the radial velocities of the fluid
to be zero which corresponds to a hydrostatic equilibrium. This condition is equivalent to
the Tolman-Oppenheimer-Volkov (TOV) equation
\[
\frac{dP(r)}{dr}=-\frac{GM(r)\rho(r)}{r^2}\left[1 +
\frac{P(r)}{\rho(r)}\right]\left[1 + \frac{4\pi r^3P(r)}{M(r)}
\right]\left[1-\frac{2GM(r)}{r}\right]^{-1}
\]
with $M(r)=4\pi \int_0^r d\xi \xi^2 \rho(\xi)$. After solving the
TOV equation we have all necessary ingredients to write down the
metric elements $A(r)$ and $B(r)$. They read \cite{Fliessbach}
\[
A(r)=\exp\left\{-2G\int_r^{\infty}\frac{d \xi}{\xi^2}\frac{M(\xi)
+ 4 \pi \xi^3 P(\xi)}{1- 2GM(\xi)/\xi}\right\},\quad
B(r)=\left[1-\frac{2GM(r)}{r}\right]^{-1}.
\]
For $r>R$ we have the Schwarzschild vacuum metric. This implies that
$\rho=P=0$ for $r \ge R$.
\subsection{Constant density}
\noindent The case of constant density is illuminating as all the formulae can be written in closed form. Indeed,
one obtains
\[
A(r)=\frac{1}{4}\left[3\sqrt{1-\frac{2r_s}{R}}-\sqrt{1-\frac{2r_sr^2}{R^3}}\right]^2\quad\mbox{and}\quad
B(r)=\left[1-\frac{2r_sr^2}{R^3}\right]^{-1}.
\]
The radial and angular velocity of a photon (being at the same time the upper bounds for
the corresponding velocities for a massive particle) are
\begin{eqnarray*}
v_r(r)&=& \frac{1}{2}\left[3\sqrt{1-2\frac{2r_s}{R}}-\sqrt{1-\frac{2r_sr^2}{R^3}}\right]\sqrt{1-\frac{2r_sr^2}{R^3}},
\nonumber \\
v_\phi(r)&=& \frac{1}{2}\left[3\sqrt{1-2\frac{2r_s}{R}}-\sqrt{1-\frac{2r_sr^2}{R^3}}\right].
\end{eqnarray*}
In Figure 4 we contrast the radial velocity of a photon moving in
matter with constant density with the velocity of a photon  moving
in matter obeying a polytropic equation of state, an issue
explained in detail below.

\subsection{Velocity of a photon moving in polytropic matter}
Let us explore a more realistic situation related to the density profile as given by
a non-relativistic hydrostatic equilibrium and a polytropic equation of state.
To be specific we use the following approximations
\begin{equation}\label{regime1}
\frac{P}{\rho}\ll 1,\quad \frac{4\pi r^3 P(r)}{M(r)}\ll 1,\quad
\frac{2GM(r)}{r}\ll 1
\end{equation}
such that the
TOV equation reduces to
\[
\frac{dP}{dr}=-G\frac{M(r)\rho(r)}{r^2}.
\]
We model matter by means of the polytropic equation of state, i.e.
$ P=K\rho^2$. This setting gives us the density as
\[
\rho(x)=\rho_0\Theta(x),\quad \Theta(x)=\frac{\sin{x}}{x},\quad
x=\sqrt{\frac{4\pi G}{2K}}r.
\]
The radius $R$ is fixed by the first zero of $\rho(x)$. This
occurs at $x_1=\pi$ and we have $R=\sqrt{2\pi K/4G}$. The
transformation linking the coordinates $x$ and $r$ can be written
as $x=\pi r/R$. The initial density $\rho_0$ is fixed by the total
mass $M$ and the radius $R$ through $\rho_0=\pi M/4R^3$. The
pressure is given through $P=K\rho^2$. From $x=\sqrt{4\pi G/2K}~r$
we get (for $r=R$) $\pi=\sqrt{4\pi G/2K}~R$ which allows to write
$K$ in terms of $R$ as follows
\[
K=\frac{2GR^2}{\pi}.
\]
Hence, the pressure can be expressed as
\[
P(r)=\frac{GM^2}{8\pi R^2}\frac{\sin^{2}{\left(\pi
r/R\right)}}{r^2}
\]
and in the interval $0\leq r\leq R$ the mass function can be rewritten
as
\[
M(r)=\frac{M}{\pi}\left[\sin{\left(\pi\frac{r}{R}\right)}-\pi\frac{r}{R}\cos{\left(\pi\frac{r}{R}\right)}\right].
\]
Concerning $A(r)$ we have
\begin{eqnarray*}
A(r)&=&{\rm
exp}\left\{-2G\int_{r}^{\infty}du~\frac{B(u)}{u^2}\left[M(u)+4\pi
u^3 P(u) \right]\right\},\\
&=&{\rm
exp}\left\{-2G\int_{r}^{R}du~\frac{B(u)}{u^2}\left[M(u)+4\pi u^3
P(u) \right]\right\}\cdot {\rm exp}
\left\{-2GM\int_{R}^{\infty}\frac{dr}{r^2\left(1-\frac{2GM}{r}\right)}\right\}.
\end{eqnarray*}
Since
\[
\int_{R}^{\infty}\frac{dr}{r^2\left(1-\frac{2GM}{r}\right)}=-\frac{1}{2GM}\ln{\left(1-\frac{2GM}{R}\right)}
\]
we obtain
\begin{eqnarray*}
A(r)&=&\left(1-\frac{2GM}{R}\right){\rm exp}
\left\{-2G\int_{r}^{R}du~\frac{B(u)}{u^2}\left[M(u)+4\pi u^3 P(u)
\right]\right\},\\
&=&\left(1-\frac{2GM}{R}\right){\rm exp}
\left\{-2G\int_{r}^{R}du~\frac{B(u)M(u)}{u^2}\left[1+\frac{4\pi
u^3 P(u)}{M(u)} \right]\right\}.
\end{eqnarray*}
Since we are in the regime (\ref{regime1}) the expression for
$A(r)$ simplifies as follows
\[
A(r) \simeq \left(1-\frac{2GM}{R}\right){\rm exp
}\left\{-2G\int_{r}^{R}du~ \frac{B(u)M(u)}{u^2}\right\}.
\]
This allows us to write the integral in the exponential as
\begin{eqnarray*}
\mathcal{I}&=&\int_{r}^{R}du~\frac{B(u)M(u)}{u^2}
=\frac{M}{\pi}\int_{r}^{R}du\frac{\sin{\left(\pi\frac{u}{R}\right)}
-\pi\frac{u}{R}\cos{\left(\pi\frac{u}{R}\right)}}
{u^2\left[1-\frac{r_s}{\pi
u}\left[\sin{\left(\pi\frac{u}{R}\right)}-\pi\frac{u}{R}\cos{\left(\pi\frac{u}{R}\right)}\right]\right]},
\\
&=&\frac{M}{R}\int_{x}^{\pi}dv\frac{\sin{v}-v\cos{v}} {v^2-\alpha
v\left(\sin{v}-v\cos{v}\right)}
\end{eqnarray*}
with $\alpha=2r_s/R$. Restricting ourselves to the case $\alpha\ll
1$ and since the function $\sin{v}-v\cos{v}$ is of bounded
variation we can approximate the denominator with $v^2$ to arrive
at
\[
A(r)=\left(1-\alpha\right){\rm
exp}\left\{-\alpha\frac{\sin{x}}{x}\right\}.
\]
Collecting our results the radial velocity of the photon becomes
\[
v_r(x)=\sqrt{(1-\alpha)\left[1-\frac{\alpha}{\pi}~(\sin{x}-x\cos{x})\right]{\rm
exp}\left\{-\alpha\frac{\sin{x}}{x}\right\}} .
\]
This velocity is plotted in Figure 4. It is a priori not obvious
that the case of radial photon velocity displays more variation
that the case of constant velocity. Note that these variations are
bounded by a local minimum and maximum and the total range of the
variation is close to one.
\begin{figure}
\begin{center}
\scalebox{.5}{\includegraphics{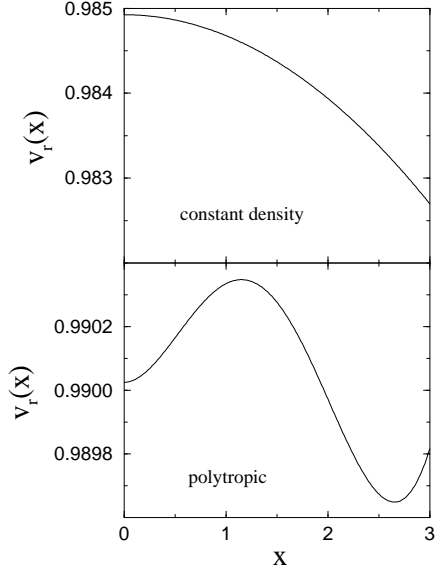}} \caption{The
radial photon's velocity in interior metrics: the upper figure is
for a constant density whereas the lower one corresponds to
polytropic configuration with polytropic index $2$. In both cases
$\alpha=2r_s/R=.0.01$.}
\end{center}
\end{figure}

\section{Conclusions}
We derived a simple expression, equation (\ref{eq:20x}), for the
photon radial velocity in dependence of the distance and as
functional of metric elements in a general static spherically
symmetric metric. Equation (\ref{eq:20x}) allows a quick
analytical insight into the behaviour of the photon radial
velocity without the necessity of solving the geodesic equation of
motion. We have applied this general result to different metrics
known from the literature. The radial velocity of massless
particles shows some interesting features:
\begin{enumerate}
\item
In the Schwarzschild case the effect on earth changes the velocity
of light to $1-1.39 \times 10^{-9}$ which is actually within the
range of the error bars of directly measured velocity of light.
Even though this makes the effect unmeasurable in laboratory, it
is good to recall that in the past the error bars were attributed
to a fuzzy definition of the meter.
\item
The Schwarzschild-de Sitter case is particularly interesting as it
might be the actual exterior space-time of spherical astrophysical
objects since a positive cosmological constant offers a simple
explanation of the acceleration of the Universe. In such a case
the radial velocity of a photon reaches a maximum value at an
astrophysical distance after which it decreases in contrast to the
Schwarzschild case where the velocity approaches one.
\item
If the central singularity of a Reissner-Nordstr\"om or
Reissner-Nordstr\"om-de Sitter black hole is not protected by a
horizon, we encounter radial velocities bigger than one. From the
point of view of Schwarzschild coordinates this is what makes the
naked singularities so special. We give
a full algebraic characterization of Reissner-Nordstr\"om-de
Sitter metric and the radial velocity of a massless particle
moving in this metric.
For instance we give conditions when the radial velocity has a local minimum and maximum
provided the charge is smaller than a critical value. We use this to define the extreme case.
We note that if the naked singularity is encountered in metrics motivated by
noncommutative geometry, the radial velocity of a photon never exceeds one.
We find this a curious fact which deserves attention.
\item
For interior metrics we contrasted the radial velocity of a photon
moving in matter of constant density with a polytropic case.
Whereas the velocity function of the constant density case is
monotonically decreasing, the polytropic case displays a local
minimum and maximum.
We mention that in both cases the variation of the radial velocity (as compared to one)
is rather small.
As a future prospect we mention that it would
be of interest to study the radial velocity of a photon (in
Schwarzschild coordinates) moving in a spherically symmetric Dark
Matter Halo.
\end{enumerate}

\end{document}